\documentclass[3p]{elsarticle}

\begin{document}

\begin{frontmatter}

\author[inst1]{Liu-Ye Meng}
\author[inst1]{Rong-Yao Yang}
\ead{ryyang@seu.edu.cn}
\author[inst1]{Wei-Zhou Jiang}
\ead{wzjiang@seu.edu.cn}

\affiliation[inst1]{organization={School of Physics, Southeast University},
             city={Nanjing},
             postcode={211189},
             country={China}}

\title{Heating anomaly of cold interfacial water under irradiation of mid-infrared pulses}

\begin{abstract}
The mid-infrared heating of interfacial water with different initial temperatures is studied using non-equilibrium molecular dynamics simulation. It is found that under the irradiation of a pulse at 3360-3380 $cm^{-1}$  the two-dimensional water monolayer on a hydrophilic surface at a lower initial temperature acquires a much larger temperature jump.  The mechanism beneath this counterintuitive phenomenon is  the enhanced transition efficiency of  the asymmetric OH stretching vibration due to  the specific oriented configuration of water molecules at lower initial temperatures. The understanding of the anomalous phenomenon clarifies  the sensitivity of  the interfacial properties of  water molecules to the temperature.
\end{abstract}


\end{frontmatter}

\section{Introduction}

The water evaporation and harvesting through interfaces play an important role in a variety of domains from the climate system, biological phenomena, to nano-devices in the aqueous environment~\cite{2021Colman-RMP,2016Zhang-small,2014kozbial}.  It is known that the interfacial properties of the materials are responsible for many interesting anomalous phenomena of water that can be seen, for instance, by the 2-dimensional (2D) ice at the room temperature under various interfacial confinements~\cite{2015Algara-Siller326,ZhuGao-265} and a water droplet formed bizarrely  on a water monolayer because of the interface beneath it~\cite{Wang2009-68}.  Moreover, there are also lots of fascinating phenomena in confined water arising from the interplay with electromagnetic fields, including the phase transition of confined 1D or 2D water to superpermeation~\cite{ZhuChang-231,ZhuChen-436}, ice-liquid phase transition of the monolayer water due to applied electric field~\cite{QiuGuo-40}, and water freezing on a pyroelectric surface during a heating-up process~\cite{EhreLavert-205}, etc.  These anomalous properties enrich the knowledge on the water complexity, as many features of the confined nano-water can be different drastically from those of bulk water.

Water molecules have the complex hydrogen-bond network which is exquisitely sensitive to temperature~\cite{Dougherty-429},  and the transient  rupture of hydrogen bonds appears to be primarily temperature dependent~\cite{MizanSavage-430}. In general, the temperature of water affects  the thermal distribution of the water molecules and the stability of  the hydrogen-bond network. The change of the temperature can have impact on both the static and dynamic characteristics of water, such as significantly faster water transport at higher temperature in nanochannels~\cite{Zhang-431}, much slower dynamics in monolayer water than those in a fully hydrated pore at ambient temperatures but the opposite at low temperature~\cite{BertrandLiu-432}, and slightly higher absorbance near 3350 $cm^{-1}$ for bulk water at a lower temperature ~\cite{LaroucheMax-351,MaxLarouche-462}. Another example is that the temperature dependence of water dynamics shows a weak crossover from fragile behavior to strong behavior in the hydration shells of elastin-like and collagen-like peptides~\cite{Vogel-433}. These interesting phenomena show that the temperature can be taken as a sensitive parameter to modulate the hydrogen bond network and affect the properties of water.

In contrast to the moderate modulation by the ambient temperature, the resonant rupture  of the hydrogen bond,  manipulated in the applied fields, can affect the water dynamics dramatically. An interesting example is that the resonant rupture of the hydrogen bond  results in the drastic  increase of the water flux in nanochannels~\cite{2013Zhang}. The resonant rupture of the hydrogen bond takes place when the energy acquisition of the water molecule through various excitations in the metastable structures cancels exactly off the hydrogen bonding. Such resonances can be intrigued by the breathing vibration mode of nanochannels~\cite{2013Zhang}, oscillations of the point charge~\cite{ZhouKou-464,kou2014electricity}, far-infrared or mid-infrared (MIR) irradiations~\cite{zhang2016fast,huang2016heating,MishraBettaque-8, yang2022orientated} and so on. In the same time, the resonant energy acquisition can result swiftly in a huge jump of the temperature of the nanosized water~\cite{zhang2016fast,huang2016heating,MishraBettaque-8}. Recently, we have worked on distinguishing the various heatings of light and heavy nano-water in the far-infrared radiation~\cite{HuangYang-6,yang2017resonant} and the energy acquisition of water systems in the MIR irradiation~\cite{yang2022orientated}. However, a study on the effect of initial temperature on the  MIR heating process of the  confined 2D water on the interface is still absent. Thus, in this work we will investigate the dependence of  the MIR heating of the interfacial water on the initial temperature and the energy acquisition induced by OH stretching vibrational resonances using molecular dynamics (MD) simulation. As the anomalous phenomena seem to be ubiquitous for aqueous processes~\cite{2016Winkel}, special attention will be paid to the new phenomenon in the energy acquisition under the MIR pulses.

\section{Computational methods}
\begin{figure}[!htb]
\centering
\includegraphics[height=8.0cm,width=8.0cm]{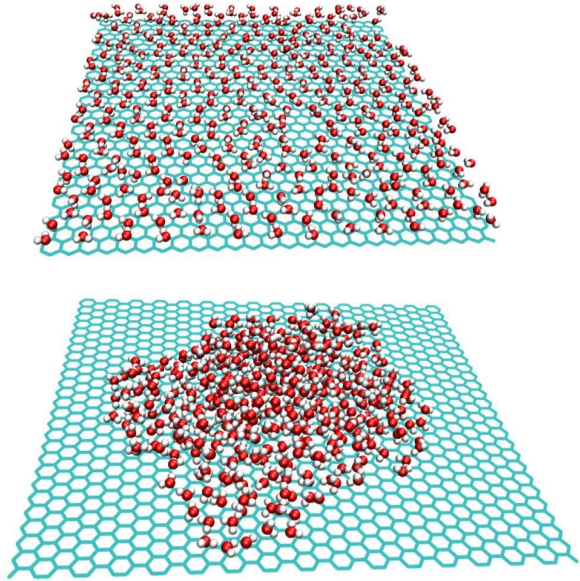}
\caption{(Color online) Snapshots of 416 water molecules on a $6.3\times6.4\ nm^2$ hexagonal surface obtained with VMD~\cite{HumphreyDalke-204}. A water monolayer (top panel) is formed on the hydrophilic surface with $q=0.6e$ charge modification and a water droplet (bottom panel) is sustained on the hydrophobic surface with $q=0.3e$ or completely without charged surface atoms. Red and white balls represent the oxygen and hydrogen atoms, respectively, and cyan lines depict the hexagonal surface whose normal is the $z$ axis.}\label{F:models}
\end{figure}

In this work, we study the initial temperature dependence of the MIR heating of water molecules on hydrophilic/hydrophobic surfaces using MD simulation. The investigated theoretical surface has a planar hexagonal structure with neighboring bond length of 0.142 $nm$ and dimensions of $6.3\times6.4\ nm^2$, which is fixed during the simulations. The Lennard-Jones parameters for the atoms on the hexagons are set to $\epsilon = -0.07\ kcal/mol,\ \sigma=0.355\ nm$. When all the surface atoms are charge neutral, it mimics  a hydrophobic surface resembling a graphene sheet, and a water droplet can be sustained on it, as shown in the bottom panel of Fig.~\ref{F:models}. On the other hand, modifying the diagonal atoms in the hexagons with the alternate positive and negative charge of the same magnitude $q=0.6e$ delineates a hydrophilic surface where a water monolayer would form, as seen in the top panel of Fig.~\ref{F:models}. Overall, this modeled hydrophilic surface is charge neutral and more details can be referred to Ref.~\cite{Wang2009-68,WangZhou-234}. The similar properties of water molecules have been verified experimentally in the  surfaces of several existing materials~\cite{WangYang-269}. As a comparison, we also examine the surface with charge modification of $q=0.3e$ which is hydrophobic, analogous to the case without any charge.
Initially, 416 water molecules on the surface are equilibrated for 2 $ns$ in a canonical ensemble with different temperatures using Langevin thermostat. Next, a consecutive 5 $ps$ long non-equilibrium simulation with explicitly applied electromagnetic field is carried out based on the equilibrated state from the previous canonical simulations.
All simulations are performed using the NAMD2 package~\cite{NAMD2020} and CHARMM27 force field~\cite{MackerellBashford-202} with periodic boundary conditions in all directions. The electrostatic interactions are handled by the particle mesh Ewald method~\cite{darden1993particle} with a real space cutoff of 1.2 nm, the same as that for the van der Waals interaction. The flexible TIP3P water model~\cite{TIP3P1983} is adopted here. The time step is 0.2 $fs$, and data are recorded every 40 $fs$.

In our simulations, the system is shone to a MIR pulse with a form of Gaussian envelope. The electric field of the pulse has its direction parallel to $z$ axis (i.e., vertical to the surface) and amplitude of $E=A\cos(2\pi c\tilde{\nu} t)e^{(t-t_c)^2/2\sigma^2}$ where $A$, $\tilde{\nu}$, $t_c$ and 2.355$\sigma$ represent in turn the maximum intensity, wavenumber, time center and full width at half maximum (FWHM) of the pulse. The concerned wavenumber is in the OH stretch region near 3300 $cm^{-1}$. Since the relevant MIR wavelength ($\sim 3\ \mu m$) is much larger than the size of the system, the applied MIR pulse is spatially uniform for all water molecules. In the following, we set $A=0.5\ V/nm$, $t_c = 2.5\ ps$, and FWHM=1.0 $ps$. The temperature is deduced from the total kinetic energy according to energy equipartition, and the temperature increment in the 5 $ps$ non-equilibrium MD simulation is defined as the difference between the average temperature in the last 1 $ps$ and that in the initial 1 $ps$.

\section{Results and discussion}

In this work, we aim to check the dependence of the energy acquisition in the MIR irradiation on the initial temperature. In the simulations, we choose $T_0=$ 270, 300 and 330 $K$ as three initial water temperatures. To simulate the various distribution of water molecules on the hexagonal surface, we add charge modifications to the diagonal atoms in the hexagons with $q=0.6e$, 0.3$e$ or 0$e$ (no charge) in various cases. We make the calculations for the system exposed to one MIR pulse in the frequency span from 3200 to 3500 $cm^{-1}$ which encompasses the frequencies of the symmetric and asymmetric OH stretch modes.

\begin{figure}[!thb]
\centering
\includegraphics[height=13cm,width=9.0cm]{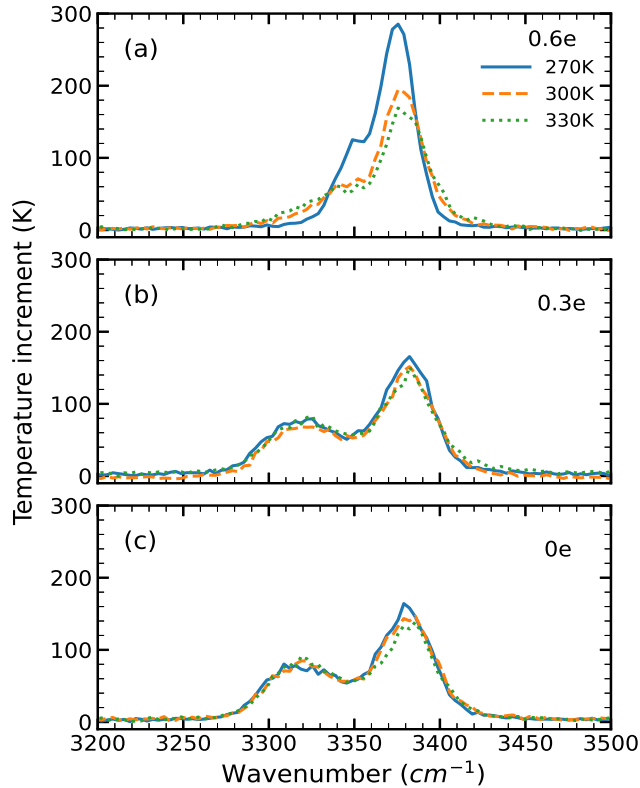}
\caption{(Color online) Temperature increments of water systems with different initial temperature and surface's charge modification as a function of the wavenumber of the MIR pulses.
} \label{F-T}
\end{figure}

We probe the kinetic and dynamical changes of the system by measuring the total kinetic energy of all atoms, which is equivalent to a temperature based on the energy equipartition. Shown in Fig.~\ref{F-T} is the temperature increment as a function of the MIR frequency with various charge modifications on the hexagonal substrate. With the increase of the electric field frequency, the acquired temperature increment of water molecules first increases and then decreases. A peak exists in the frequency profile of the temperature for all cases. The peak is a result of the resonant energy absorption under the MIR irradiation. The metastable potential of the water molecule, responsible for this resonant absorption, is given by the intramolecular OH stretching vibrations.

As shown in Fig.~\ref{F-T}a where the charge modification is 0.6e, the maximum temperature increment of water molecules appears around 3375 $cm^{-1}$ which is the resonant frequency of the asymmetric OH stretching vibration in the potential of the water model TIP3P~\cite{yang2022orientated,JansenCringus-428}. In principle, there is another resonant peak at the frequency ($\sim3 320\ cm^{-1}$) of the symmetric OH stretching vibration. It is, however, almost fully suppressed   with the MIR electric field vertical to the surface, since the transition dipole moment of the symmetric OH stretching vibration is mostly parallel to the surface in this case.  The maximum temperature increments are 285, 194 and 169 $K$ for the interfacial water monolayer with the initial temperature of 270, 300 and 330 $K$, respectively. Obviously, this is an anomalous  phenomenon in contrast to the usual temperature climbing under the same heating and  thermal conduction.

For a more comprehensive look at this anomalous phenomenon, it is necessary to check the dependence on the charge modification. Shown in Fig.~\ref{F-T}b-\ref{F-T}c are the temperature increments with the surface modified by 0.3$e$ or no charge (0$e$).  The energy absorption from the MIR pulse gives rise to similar curves but with a lower temperature increment than that in Fig.~\ref{F-T}a near 3375 $cm^{-1}$.  The maximum temperature increments for the case with initial temperature $T_0=270\ K$ is slightly large than the results with  $T_0=300,\ 330\ K$. This tendency is consistent with experimental measurements on bulk water~\cite{LaroucheMax-351,MaxLarouche-462}. Noticeably, the peak owing to excitation of symmetric OH stretch shows up near 3320 $cm^{-1}$ as a result of more uniform distribution of water molecular orientation in these cases.

\begin{figure}[!thb]
\centering
\includegraphics[height=6.5cm,width=9.0cm]{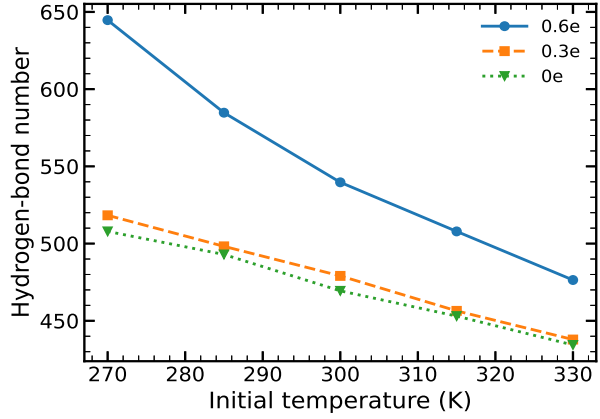}
\caption{(Color online) The number of hydrogen bonds as a function of equilibrium temperature for surfaces with different charge modifications.} \label{F-Hbond}
\end{figure}
On the other hand, the hydrogen bond is much relevant, since more hydrogen bonds can keep water molecules on the surface  better ordered. According to the criterion of the oxygen distance less than 3.5 {\AA} and hydrogen-bond angle $\leq30^\circ$, we count the number of hydrogen bonds of water molecules using the data from the equilibrium simulations, which is shown in Fig.~\ref{F-Hbond}. For the temperature  much lower than the boiling point, water molecules are well bound by hydrogen bonds, while the number of hydrogen bonds decreases as the temperature increases, because higher temperature is more conducive to destroy the hydrogen bond network. In addition to the temperature effect, the charge modification can make difference for the hydrogen bonds, as shown in Fig.~\ref{F-Hbond}. The charge modification of 0.6$e$ can well fix the water molecules on the surface with more hydrogen bonds. With  the charge modification of 0.3$e$ or less, the water molecules will stack randomly up to several layers with a rather amorphous surface on which molecules are less connected by the hydrogen bonds. Roughly to say, less hydrogen bonds of water molecules mean to be less oriented, which is injurious to the transition to the OH stretching vibration modes in an linearly polarized MIR pulse.

\begin{figure}[!thb]
\centering
\includegraphics[height=9cm,width=9.0cm]{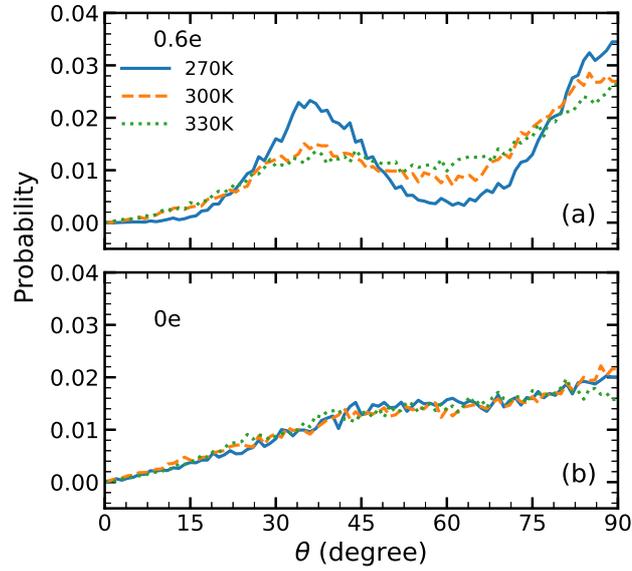}
\caption{(Color online) Angular distributions of the water molecules at equilibrium with different temperatures and charge modifications. $\theta$ is defined as the acute angle between the line connecting two hydrogen atoms in a water molecule and the surface normal (i.e., $z$ axis).} \label{F-ang}
\end{figure}

More quantitatively, the anomalous energy acquisition can be understood by the transition probability of the OH stretching vibrations.  The non-linear water molecule has symmetric and asymmetric OH stretching modes. It is known that the energy absorption intensity of the water molecule  in the MIR pulse  depends on the transition probability which is proportional to $\cos^2\theta$, where $\theta$ is the angle between the direction of the MIR electric field  and the transition dipole moment~\cite{yang2022orientated,wout1997femto}. Therefore, the smaller value of average $\theta$ is, the greater the absorption intensity will be. The total energy acquisition of the system can be obtained from the integration of the angular distribution over the energy absorption of the individual water molecules. Shown in Fig.~\ref{F-ang} are the angular distributions of the water molecules.  With surface charge modification of $q=0.6e$, we can see from Fig.~\ref{F-ang}a that the peaks at a smaller angle $\theta\approx 40^\circ$ distinguish the distinct proportion of water molecules, which is much higher  at $T_0= 270 K$ than at $T_0= 330 K$. Thus, the higher water proportion distributed at small angles with more efficient energy absorption accounts successfully for the anomalously higher temperature increment for the system at a lower initial temperature. Without charge modification or with $q=0.3e$ (not shown in Fig.~\ref{F-ang} due to its similarity with $q=0e$), the peak of the angular distribution at small angles becomes obscure, indicating that orientations of water molecules are distributed more uniformly in space. The attractive interaction from the surface and the interlocking effect from the hydrogen-bond network are largely weakened in these two cases, which make it difficult for water molecules to maintain a orientation-preferential morphosis. This explains its considerably diminished dependence of the MIR absorption on the temperature.

\section{Conclusion}

We have investigated the heating of interfacial water with different initial temperature under the irradiation of the MIR pulse using MD simulation. A prompt heating appears at the frequency ranging from 3300 to 3400 $cm^{-1}$ due to the excitation of OH stretching vibration of water molecules. Anomalously, the water monolayer on a hydrophilic surface with a colder initial state acquires more energy from the MIR pulse with a substantially larger temperature escalation, while a transition of the interfacial 2D distribution of water to amorphously hydrophobic stacking on the surface leads to  a clear attenuation of the anomalous heating.  This interesting phenomenon on the hydrophilic surface results from the enhanced excitation of asymmetric OH stretch subject to more oriented configuration of water molecules under lower temperature with the help from surface attraction and hydrogen-bond interlocking. Since interfacial water is ubiquitous in nature, these findings that reveal the sensitive temperature dependence of the MIR resonant energy acquisition may have practical applications in various systems in the aqueous environment.

\section*{acknowledgement}
The work was supported in part by the National Natural Science Foundation of China under Grant No. 11775049, the China Postdoctoral Science Foundation under Grant No. 2021M690627, and the Fundamental Research Funds for the Central Universities under Grant No. 2242022R20041.

\bibliographystyle{elsarticle-num}

\bibliography{ref-heat}

\end{document}